\documentclass[aps,pre,reprint,amssymb,superscriptaddress,showpacs,showkeys,
]{revtex4-1}

\usepackage{color}

\usepackage{graphicx}% Include figure files
\usepackage{dcolumn}% Align table columns on decimal point
\usepackage{bm}% bold math
\usepackage{epsf}

\begin{document}

\preprint{APS/123-QED}

\title{Sensitivity of asymmetric rate-dependent critical systems to initial conditions: insights into cellular decision making}% Force line breaks with \\

\author{Nuno R. Nen\'{e}}
\affiliation{Department of Genetics, University of Cambridge, CB2 3EH Cambridge,  UK}%

\author{James Rivington}
\affiliation{Department of Mathematics,
University College London, Gower Street,  WC1E 6BT London, UK}%

\author{Alexey Zaikin}
\affiliation{Department of Mathematics,
University College London, Gower Street,  WC1E 6BT London, UK}
\affiliation{Institute for Women's Health,
University College London, Gower Street,  WC1E 6BT London, UK}%
\affiliation{Lobachevsky State University of Nizhny Novgorod, Nizhny Novgorod, Russia}

\date{\today}% It is always \today, today,

\begin{abstract}

The work reported here aims to address the effects of time-dependent parameters and stochasticity on decision-making in biological systems. We achieve this by extending previous studies that resorted to simple normal forms. Yet, we focus primarily on the issue of the system's sensitivity to initial conditions in the presence of different noise distributions. In addition, we assess the impact of two-way sweeping through the critical region of a canonical Pitchfork bifurcation with a constant external asymmetry. The parallel with decision-making in bio-circuits is performed on this simple system since it is equivalent in its available states and dynamics to more complex genetic circuits. Overall, we verify that rate-dependent effects are specific to particular initial conditions. Information processing for each starting state is affected by the balance between sweeping speed through critical regions, and the type of fluctuations added. For a heavy-tail noise, forward-reverse dynamic bifurcations are more efficient in processing the information contained in external signals, when compared to the system relying on escape dynamics, if it starts at an attractor not favoured by the asymmetry and, in conjunction, if the sweeping amplitude is large.

\end{abstract}

\pacs{87.18.-h, 87.18.Cf, 87.16.Yc, 05.45.-a,87.18.Tt}% PACS, the Physics and Astronomy
\keywords{Pitchfork bifurcation, rate-dependent effects on attractor selection, Gaussian noise, L\'{e}vy noise, cellular decision making.}%Use showkeys class option if keyword
\maketitle

%\tableofcontents

\section{\label{sec:Intro}Introduction}

The fidelity with which cellular systems respond to external fluctuations has generated an increasing interest in quantifying the resultant downstream effects that elicit dynamic responses \cite{Shahrezaei2008,Bowsher2012,Bowsher2013,Dattani2016}. The idea of robustness in the face of unpredictable external drivers has also been prevalent in other areas, particularly in evolutionary biology. There, systems are seen, to an extent, as being the result of continuously changing environments determining fitness \cite{Mustonen2008,Mustonen2009}. In more clinically oriented applications, albeit in the realm of evolutionary biology, the idea of an external control has also been important in the design of adaptive and optimal therapies under a stochastic control paradigm \cite{Gatenby2009,Fischer2015}. Adding to this body of work, recent developments in bio-pattern formation have shown that path-dependent effects imposed by external sources are a significant component of observed phenotypic outcomes \cite{Palau-Ortin2015}. The subject of an external driver inducing bifurcations in the underlying intrinsic dynamics has been less debated in biology. The study of such systems opens up several research avenues that only recently have  attracted considerable interest \cite{Huang2007,Nene2012,Nene2012b}. Therefore, there is scope for extensive testing, from a computational point of view, of the relevant features brought from stochastic open-systems undergoing critical transitions \cite{Cross1993,Berglund2006,Hobbs2013,Ashcroft2013,Perryman2014}. The main ingredients from this area that we will explore in the context of decision making in biology are the following: critical parameter time-dependence; passage through a critical region at different rates; stochasticity hindering the convergence to any of the new emerging states.

% ----------------------
% ----------------------
\begin{figure*}[!tb]
\includegraphics[width=0.29\textwidth]{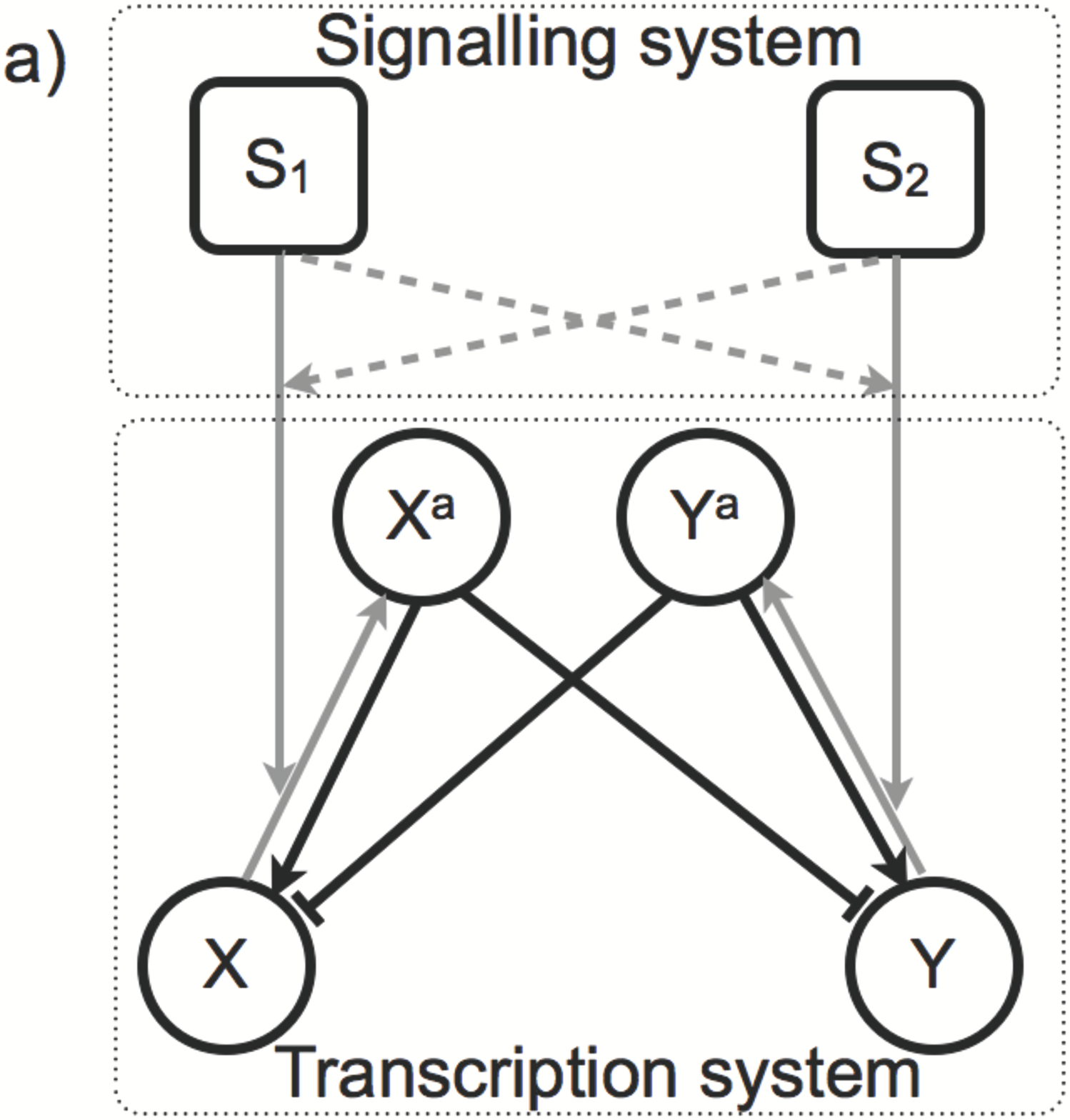}
\includegraphics[width=0.355\textwidth]{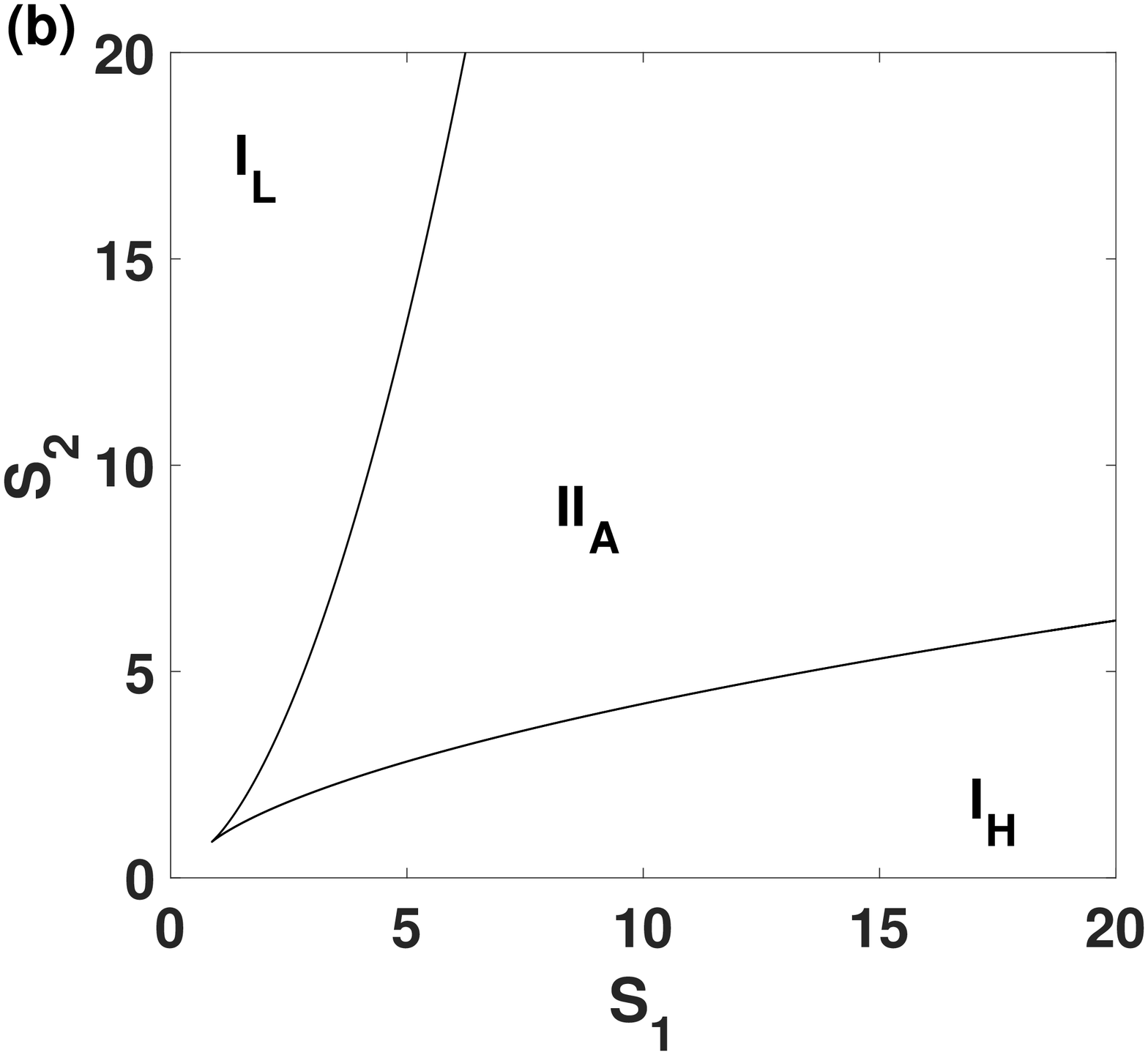}
\caption{Representative low order genetic switch with external stimulation. (a) Schematic representation of circuit: nodes $X$ and $Y$ stand for transcription factors that can be activated to generate $X^{a,b}$. $S_{1,2}$ represents external or upstream signals inducing the activation; black lines represent transcriptional reactions, leading to induction or repression of production of $X$ and $Y$; grey lines depict activation or protein-protein interactions. (b) Phase diagram in the space of $(S_{1},S_{2})$. Thin lines represent borders between different regimes:  $I_{L,H}$ stands for monostability, with $X$ having a low ($L$) or a high value ($H$).  $II_{A}$ denotes bistability between two states at which $X$ and $Y$ have opposite concentrations, (high, low) or (low, high). See \cite{Nene2012} for details of the underlying equations.}\label{fig:Fig_BioNet}
\end{figure*}
% ----------------------
% ----------------------

In this work, we extend previous studies that sought applications in network biology and that were developed by some of the authors \cite{Nene2012,Nene2012b,Nene2013,Alagha2013}. A representative low order circuit underlying such studies, the integrative signalling-gene regulatory switch, is depicted in Fig.~\ref{fig:Fig_BioNet} (a); its structure can be tweaked so as to resemble other circuits behind observed phenomena (see for example \cite{Pfeuty2009}). In addition, in the same figure, we also show the phase diagram corresponding to a set of non-linear differential equations including activation, translation and transcription of crucial proteins, in this case transcription factors. By varying the values of signals $S_{1}$ and $S_{2}$, which in the case of Fig.~\ref{fig:Fig_BioNet} (a) work as the external drivers,  we are able to generate typical regimes observed in systems relevant to experiments \cite{Huang2007,Enver2009,Schroter2015}. The mechanism of Speed-dependent Cellular Decision Making (SdCDM) \cite{Nene2012}, which arises from crossing the critical region ($I_{L,H}$ to $II_{A}$) at different rates, is one of such regimes. Here, instead of relying on the integrative genetic switch once again, we will opt for simple standard norm forms that exhibit similar behaviours and regimes to those represented in  Fig.~\ref{fig:Fig_BioNet} (b). In fact, the effects of $S_{1}$ and $S_{2}$ can be economically modelled by a supercritical Pitchfork bifurcation normal form with coupled time-dependent critical parameter and external asymmetry (see Eq.~\ref{eq:PitchFork}) \cite{Nene2013}. This normal form has been successfully used in the study of genetic circuits behind, for example, decision making in haematopoietic cell differentiation regulated by $GATA1$ and $PU.1$ \cite{Huang2007,Enver2009}. More recent work has also resorted to the idea of the bi-stable potential with external drivers in order understand the influence of signalling on expression dynamics in the $GATA-NANOG$ circuit in embryonic stem cells \cite{Schroter2015}. Schroter and co-workers \cite{Schroter2015} postulated that the integrated system and respective model might be a general network architecture to integrate the activity of signal transduction pathways and transcriptional regulators and, in this biological context, serve to balance proportions of cell fates in several environments. The idea of cellular response as integrated response to external drivers is, therefore, also present in this work. Despite the issue of bifurcations, or critical transitions, not making part of the model underlying the study reported in \cite{Schroter2015}, their system can also be tested under the framework highlighted below and  explored in \cite{Nene2012,Nene2013}. 

Our choice of a standard normal form allows us to link our findings to previous theoretical work on dynamic bifurcations and, ultimately, serves as a bridge to investigations of the importance of rate-dependent effects in complex noisy genetic networks. The work presented here is, above all, an investigation into the sensitivity to initial conditions when all of the ingredients reported above are present. Unlike before \cite{Nene2013}, we study the effects of both forward and reverse bifurcations when trajectories start in the bi-stability region. We further delve into the importance of fluctuations following different distributions: the typical Gaussian and that arising in the literature of L\'{e}vy processes in biology \cite{Xu2016}. The latter is an important alternative to modelling transitions between states even when noise amplitudes are small and constitutes a viable candidate for modelling cell fate decision as an escape problem \cite{Liu2004,Jaruszewicz2013,Xu2016}.

\section{Dynamically bifurcating systems with noise and asymmetries}

\subsection{Forward bifurcations}\label{sec:FwdBif}

% ----------------------
% ----------------------
\begin{figure*}[!tb]
\includegraphics[width=1\textwidth]{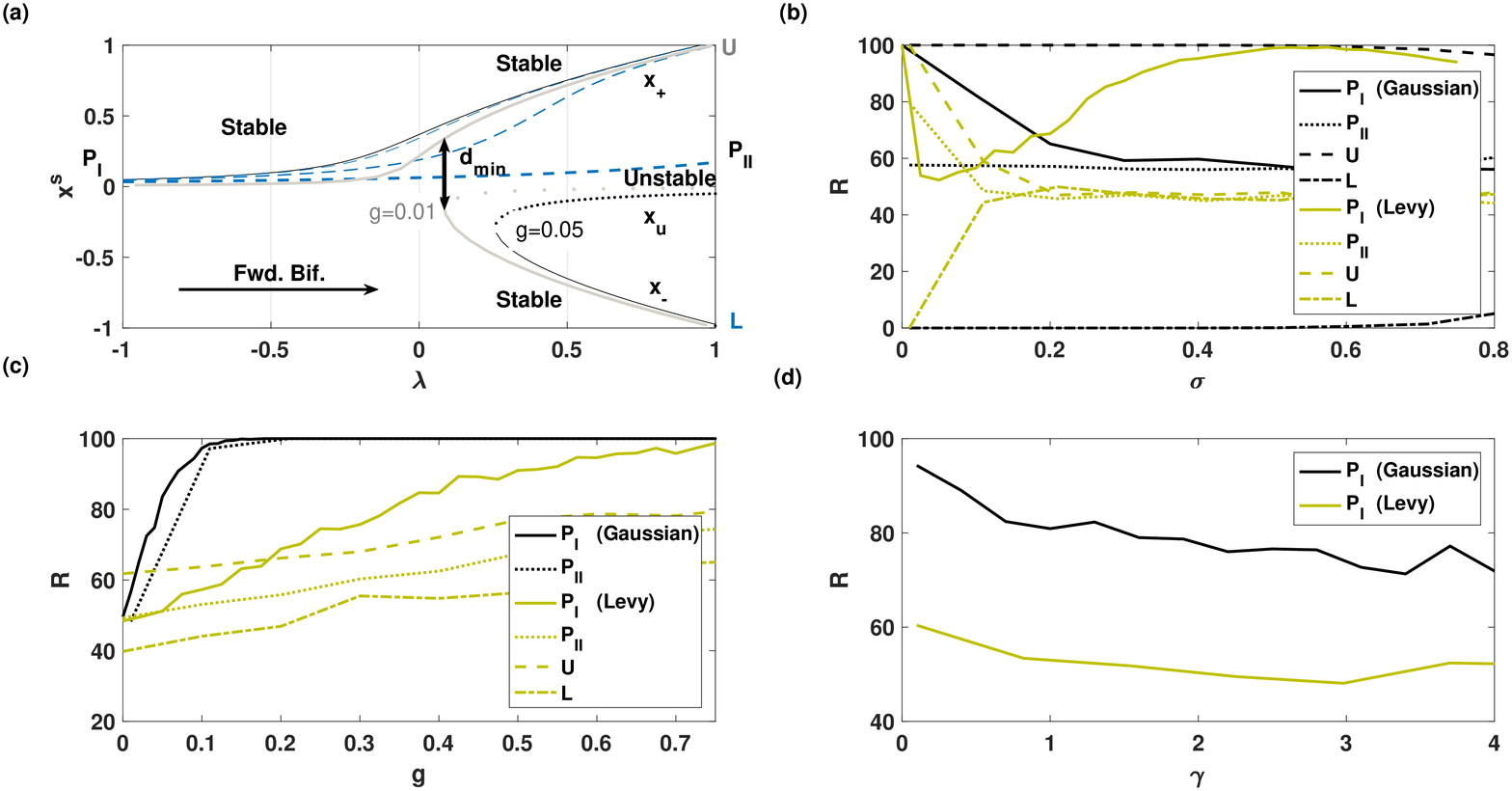}
\caption{Rate-dependent effects in forward dynamic bifurcations in conjunction with asymmetries and noise. (a) Bifurcation diagrams for $g=0.01$ and $g=0.05$. Also shown in blue are the deterministic trajectories for $g=0.05$ when $\gamma=0.01$ (thinner), $0.1$ (intermediate) and $1$ (thicker) and $x(0)=x^{s}(\lambda=-1)$. (b) Selectivity $R$ with noise amplitude $\sigma$, for $g=0.05$ and $\gamma=1$, when the system starts at $x^{s}(\lambda=-1)$, represented as $P_{I}$. Also shown is the selectivity for the system when $\lambda(t)=1$ and the initial conditions are at the upper branch ($U$), lower branch ($L$), and at the point equal to the steady-state $x^{s}(\lambda=-1)$  ($P_{II}$). (c) With asymmetry $g$, for $\sigma=0.1$ and $\gamma=1$. As in (c) we also show results for the case where $\lambda$ is not swept and remains equal to 1. Curves for starting points $U$ and $L$ did not change with $g$ and are not shown. (d) With sweeping speed $\gamma$, when $g=0.05$ and $\sigma=0.1$.  Black lines: Gaussian noise. Green: L\'{e}vy noise, with $\mu = −1.8$, $c = 1/100$ and $\xi^{u} = 5$ (see Eq.~\ref{eq:LevyCDF} and \ref{eq:LevyDensity}). 1000 trajectories were used in the calculation of $R$, the percentage attracted to $x_{+}$.}\label{fig:Fig_FwdBifGaussianLevy}
\end{figure*}
% ----------------------
% ----------------------

A typical bifurcation representing decision-making in biology \cite{Nene2012, Huang2007} or second-order phase transitions in physical systems \cite{Cross1993,Moss1985,Kondepudi1983,Nicolis2000,Nicolis2005,Nicolis2005a} is that underlying Eq.~(\ref{eq:PitchFork}). In the case where the external asymmetry $g(t)$ is zero and the bifurcation parameter $\lambda$ is independent of time, Eq.~(\ref{eq:PitchFork}), which represents a supercritical Pitchfork normal form, has the unique asymptotically stable solution $x^{s}=0$ when $\lambda<0$. For positive values of  $\lambda$, three solutions can be clearly shown to appear: the asymptotically stable branches given by $\pm\sqrt{x}$ and the trivial unstable solution $x^{s}=0$. 

\begin{eqnarray}
&\dot{x}=\lambda (t) x-x^{3}+g (t) \label{eq:PitchFork}\\
&\lambda(t)=\lambda_{0}+\gamma t \label{eq:Lambda}
\end{eqnarray}

In the work presented here, we are interested in the solutions of Eq.~(\ref{eq:PitchFork}) when $g(t)$ is not zero. This asymmetry can be seen as a representation of discrepancies between upstream signals to a circuit regulating cell fate decision (see for example Fig.~\ref{fig:Fig_BioNet} (a) and also \cite{Nene2013}). If the asymmetry is held at a constant value $g$, the previous bifurcation point disappears and a new picture emerges made of 3 branches at $\lambda=\lambda_{c}$: a connected set of solutions with positive values, $x_{+}$, a disconnected branch with negative values, $x_{-}$, and an unstable branch $x_{u}$ (see Fig.~\ref{fig:Fig_FwdBifGaussianLevy} (a)). In this imperfect bifurcation the branches are separated by a minimum distance $d_{min}=(\Delta X)_{\lambda = \lambda_{c}}=(x_{+}-x_{-})_{\lambda = \lambda_{c}}=3\left(\frac{g}{2}\right)^{\frac{1}{3}}$; as $g$ is increased so is the distance between solutions at $\lambda_{c}$. In addition, the critical point $\lambda_{c}$ is displaced towards positive values of $\lambda$ by $3\left(\frac{g}{2}\right)^{\frac{2}{3}}$ \cite{Nene2013}. We should add that the connected and disconnected branches invert their positions if, contrary to Fig.~\ref{fig:Fig_FwdBifGaussianLevy} (a), we impose a negative asymmetry. Other types of bifurcation have been selected as models of cell decision making in biology \cite{Huang2007,Guantes2008,Enver2009,Chickarmane2009,Pfeuty2009}. The idea behind the work presented here is still valid in those cases although the framework has to be adapted for optimal representation.

As was explored in the previous work \cite{Nene2013}, our aim is to understand the effects of drivers on the behaviour of a system regulated by Eq.~(\ref{eq:PitchFork}). In \cite{Nene2013}, we studied the effect of a ramped bifurcation parameter (Eq.~\ref{eq:Lambda}) and a coupled time-dependent transient asymmetry $g(t)$. This work was motivated by integrative signalling-gene regulatory circuits (see Fig.~\ref{fig:Fig_BioNet}) that exhibit the same critical behaviour \cite{Nene2012}. By studying the effect of sweeping the system through the critical region under different $\gamma$ rates (Eq.~\ref{eq:Lambda}), in the presence of fluctuations, we were able to prove the existence of speed-dependent effects in branch selectivity. The same holds for constant external asymmetries, even if they are much smaller than 1 \cite{Moss1985,Kondepudi1983}. In both cases, the percentage of trajectories ($R$, selectivity) in a stochastic simulation that are attracted to the branch favoured by the asymmetry is proportional to Eq.~\ref{eq:SelAnalytica}, where $\sigma$ represents the amplitude of fluctuations, $g$ the asymmetry, $\gamma$ the critical parameter sweeping speed (Eq.~\ref{eq:Lambda}), $erf(.)$ the error function and $\alpha=0$ and $\beta=1$ if $g$ is constant. Two of the main contributors to the sensitivity of the system to the effect of the external asymmetry is the inflexion of the connected branch and the position of $\lambda_c$ (observe Fig.~\ref{fig:Fig_FwdBifGaussianLevy} (a)) \cite{Kondepudi1983,Moss1985,Erneux1991,Nene2013}. These factors, in conjunction with lower sweeping speeds, increase branch selectivity in a significant way due to lower switching delays \cite{Kondepudi1983,Moss1985,Erneux1991,Berglund2006,Nene2013}; these are defined as the additional time spent near the potential unstable boundary after the system goes through the critical point \cite{Berglund2006}. This result is observed even if the amplitude of fluctuations with respect to the asymmetry is large \cite{Nene2013}.

\begin{eqnarray}
R \propto \frac{1}{2}\left[1+erf\left[\frac{g}{\sigma}\left( \alpha +\beta \left(\frac{\pi}{\gamma}\right)^{\frac{1}{4}}\right)\right]\right] \times 100 \label{eq:SelAnalytica}
\end{eqnarray}

The switching delay dependence on sweeping speed can be clearly verified in Fig.~\ref{fig:Fig_FwdBifGaussianLevy} (a), where several deterministic trajectories are plotted for a constant $g=0.05$. The system was initially started at the stable branch for $\lambda=-1$ and  $\lambda$ was subsequently changed according to the linear law represented in Eq.~\ref{eq:Lambda}. As is evident, lower sweeping rates induce paths that are further away from the unstable state when the critical region emerges. The instant where the switch begins can be demonstrated to be proportional to $1/\gamma$, a factor that also influences the probability of reaching the branch favoured by $g$ when fluctuations are incorporated \cite{Moss1985,Kondepudi1986a,Nicolis2000,Grossmann1990,Nene2013}. The effects of $g$ also reduce the probability of escape over the potential barrier, located along the unstable state,  and which can be estimated to be located at $-\frac{g}{\lambda-\lambda_{c}}$ far beyond $\lambda_{c}$. The time-scales and probability for these transitions can be modelled under the Kramer's classical theory \cite{Gardiner1994}.

\subsubsection{The effect of noise model on branch selectivity}\label{sec:StochFluct}

One of the motivations for the work presented here is understanding if differences in noise distribution affect differently the system's memory of initial positions. In addition, we also aim to verify once again if rate-dependent effects are still a determinant in state selection. Our previous work tested memory of transient signals when the system was driven through the critical region in one direction only \cite{Nene2012,Nene2013}.  Here, we change slightly the scope and invest in simulations that highlight both the effect of initial conditions and a forward-reverse bifurcation scenario. This is closer to the situations observed in experimental biology where signals often have a transient character \cite{Werner2005} or more complex profiles \cite{Behar2010,Dattani2016}.

Before evaluating the forward-reverse dynamic bifurcation scenario, let us first address the simple system represented in Fig.~\ref{fig:Fig_FwdBifGaussianLevy} (a) when fluctuations are present ($\xi (t)$ in Eq.~\ref{eq:EqWithNoise}), so that we pin down the crucial aspects underlying branch selectivity for the noise distributions tested here.

\begin{eqnarray}
\dot{x} =&\lambda (t) x-x^{3}+g+\xi (t) \label{eq:EqWithNoise}
\end{eqnarray}

As expected from the diagram represented in Fig.~\ref{fig:Fig_FwdBifGaussianLevy} (a) and the deterministic trajectories plotted in blue, the overall shape of the distribution of trajectories when the control parameter $\lambda$ is passed through the critical region is approximately Gaussian; at the same time it gradually drifts due to the positive external asymmetry $g$. Along with this bias in the process, the distribution also spreads up to the point where the critical value is reached $\lambda=\lambda_{c}=3\left(\frac{g}{2}\right)^{\frac{2}{3}}$; at this moment it starts reflecting the bi-modality exerted by the bi-stability region \cite{Nene2013}. Around the critical region and just before the onset of bi-stability, fluctuations are amplified and the convergence times towards the attractor are hindered. This may be counterbalanced by a strong eternal field in conjunction with a slowly changed $\lambda$ (Fig.~\ref{fig:Fig_FwdBifGaussianLevy} (c) and (d)) \cite{Moss1985,Grossmann1990,Nene2013}. 

Two distributions were tested for the noise term $\xi (t)$ in Eq.~\ref{eq:EqWithNoise}: the standard Gaussian and the L\'{e}vy distribution. The assumption of a Gaussian is consistent with previous work \cite{Moss1985,Kondepudi1983,Nicolis2005a,Nene2013} and follows the typical assumptions in the literature: zero mean and correlation $\left\langle \xi (t),\xi (t') \right\rangle=\sigma^{2} dt \delta(t-t')$. The L\'{e}vy noise term is, on the other hand, less common. Its usage in biology was recently proven to be a valid approach to studying the effect of fluctuations in bi-stable systems \cite{Xu2016}. We resort to this additional noise paradigm with the intent of understanding if the long tail characteristic of the L\'{e}vy distributed noise influences considerably the memory of initial conditions. This follows from the work on the role of stochasticity in biology as a major determinant of cell decision outcomes in different environments, by way of crossing/escaping over potential barriers \cite{Liu2004,Jaruszewicz2013,Xu2016} or by noise-induced symmetry breaking \cite{Kobayashi2011}. 

In order to test the L\'{e}vy noise it was necessary to truncate the distribution at an upper level, thus avoiding impractical extreme values. The percentile function for a L\'{e}vy distribution truncated to the support $\xi \in [\mu, \xi^{u}]$, where $\mu$ is the normal lower truncation due to the shift parameter $\mu$ and $\xi^{u}$ is the upper truncation level, can be observed in Eq.~\ref{eq:LevyDensity}, where $F(\xi)$ is the L\'{e}vy cumulative density function (Eq.~\ref{eq:LevyCDF}).

\begin{eqnarray}
p(\xi)=&\frac{c}{2 [erfc^{-1}(F(\xi^{u})\xi)]^{2}}+\mu \label{eq:LevyDensity}\\
F(\xi)=&erfc(\sqrt{\frac{c}{2(\xi-\mu)}})\label{eq:LevyCDF}
\end{eqnarray}

Here, $c$ is the scale parameter of the L\'{e}vy distribution and $erfc^{-1}(x)$ the inverse complementary error function.

Varying the intensity parameter, $\sigma$, for the L\'{e}vy distributed noise model  (with
$g = 0.05$, $\gamma = 1$), we can observe that for $\sigma  \leq 0.08$ it follows a similar behaviour to the Gaussian distributed model, where
the percentage of paths attracted to each attractor converges towards equality as the amplitude increases (Fig.~\ref{fig:Fig_FwdBifGaussianLevy} (b)). Contrary to the Gaussian model, for $\sigma >  0.08$, the probability of reaching the  attractor favoured by the asymmetry then converges towards 1. To gain an understanding of the general path behaviour leading to the results discussed here, we have to recall that when crossing the $\sigma$ threshold observed in Fig.~\ref{fig:Fig_FwdBifGaussianLevy} (b), a qualitatively different regime ensues. Beyond $\lambda_{c}$ branch to branch transitions can occur that hinder the identification of the signal represented by $g$. Since the propensity for transitions to take place is larger with L\'{e}vy noise, the percentage of trajectories reaching $x_{+}$ should further decrease in a much more significant way. Nevertheless, there are two fundamental components at play. First, the escape rate diminishes as $\lambda$ reaches higher values, especially from $x_{+}$ to $x_{-}$; this arises from the difference between the potential associated with $x_{+,-}$ and $x_{u}$ becoming larger as $\lambda$ is swept \cite{Nicolis2005a}. Therefore, the potential difference traps the system in $x_{+}$ due to $g$. On another side, if the simulations are long enough, the chances of converging towards the positive branch are higher due to the positive heavy-tailed  L\'{e}vy noise term. Consequently, for sufficiently large noise amplitudes and longer trajectories, most paths eventually converge to the positive attractor as the synergy between the two components emphasized above is stronger than the destructive power of fluctuations. This explains the unusual curve in Fig.~\ref{fig:Fig_FwdBifGaussianLevy} (b), when the initial condition is at $P_{I}$ and $\lambda$ is driven through the critical region.

The effect of the asymmetry $g$ as a state selector can be visualized in Fig.~\ref{fig:Fig_FwdBifGaussianLevy} (c). For both noise distributions the capture of the trajectories by the upper branch, for a constant sweeping rate $\gamma=1$, is more efficient for higher values of $g$. As mentioned above, this results from both the position of the critical value $\lambda_{c}$ and the inflexion of the upper branch (see also Eq.~\ref{eq:SelAnalytica} for an approximate expression). This result had been seen in previous publications \cite{Nene2012,Nene2013} and follows intuitively from the observation of the deterministic trajectories depicted in Fig.~\ref{fig:Fig_FwdBifGaussianLevy}. For these specific results, the noise amplitude used is below the threshold mentioned in the previous paragraph. The evaluation of the impact of the external asymmetry as a state selector is, therefore, not confounded with the effects of the positive heavy-tail of the L\'{e}vy distribution (Fig.~\ref{fig:Fig_FwdBifGaussianLevy} (b)).

Cell decision making has been widely modelled as a process where the most probable outcome is already encoded in the distribution of attractors; this perspective sees the desired decision outcomes arising simply by attractor to attractor transitions induced by noise (see for example \cite{Xu2016}). Here, as was the case of previous publications by some of the authors of this study, we evaluate a different decision-making paradigm. Nevertheless, it is important to verify which scenario is more efficient in processing information. In Fig.~\ref{fig:Fig_FwdBifGaussianLevy} (b) and (c), the values of $R$ computed when $\lambda$ is time-dependent and follows Eq.~\ref{eq:Lambda} can be compared with those when it is held at its maximum. In the latter, branch selectivity is solely determined by escape dynamics, not the dynamic bifurcation. Overall, when $\lambda$ is held at 1, the Gaussian term requires much larger noise amplitudes to tilt the percentage towards $50$ and is, therefore, ineffective in eliciting jumps over the potential barrier. The L\'{e}vy distributed noise allows, on the other hand, for jumps to occur across the potential barrier which explains the tendency for $R$ to reach $50\%$ at much lower noise amplitudes. Comparing the selectivity obtained under a dynamic bifurcation (starting point $P_{I}$ in Fig.~\ref{fig:Fig_FwdBifGaussianLevy} (b) and (c)), with a comparable situation resulting from escape over the potential barrier (starting point $P_{II}$ in Fig.~\ref{fig:Fig_FwdBifGaussianLevy} (b) and (c)), we verify that, over most $\sigma$'s and asymmetries, crossing through the critical region enhances selectivity. This is a fundamental result for understanding the results in the forward-reverse dynamic bifurcation explored in section~\ref{sec:FwdRevBif}.

Regarding the rate-dependent effects on the propensity for reaching the attractors favoured by $g$, it is clear that this state selection mechanism is present when both noise distributions are used (Fig.~\ref{fig:Fig_FwdBifGaussianLevy} (b)). As observed in previous studies \cite{Nene2013}, larger $\gamma$'s destroy the information contained in $g$, a consequence felt stronger if the heavy-tale noise distribution is imposed.

\subsection{Forward-reverse bifurcations}\label{sec:FwdRevBif}

% ----------------------
% ----------------------
\begin{figure*}[!tb]
\includegraphics[width=1\textwidth]{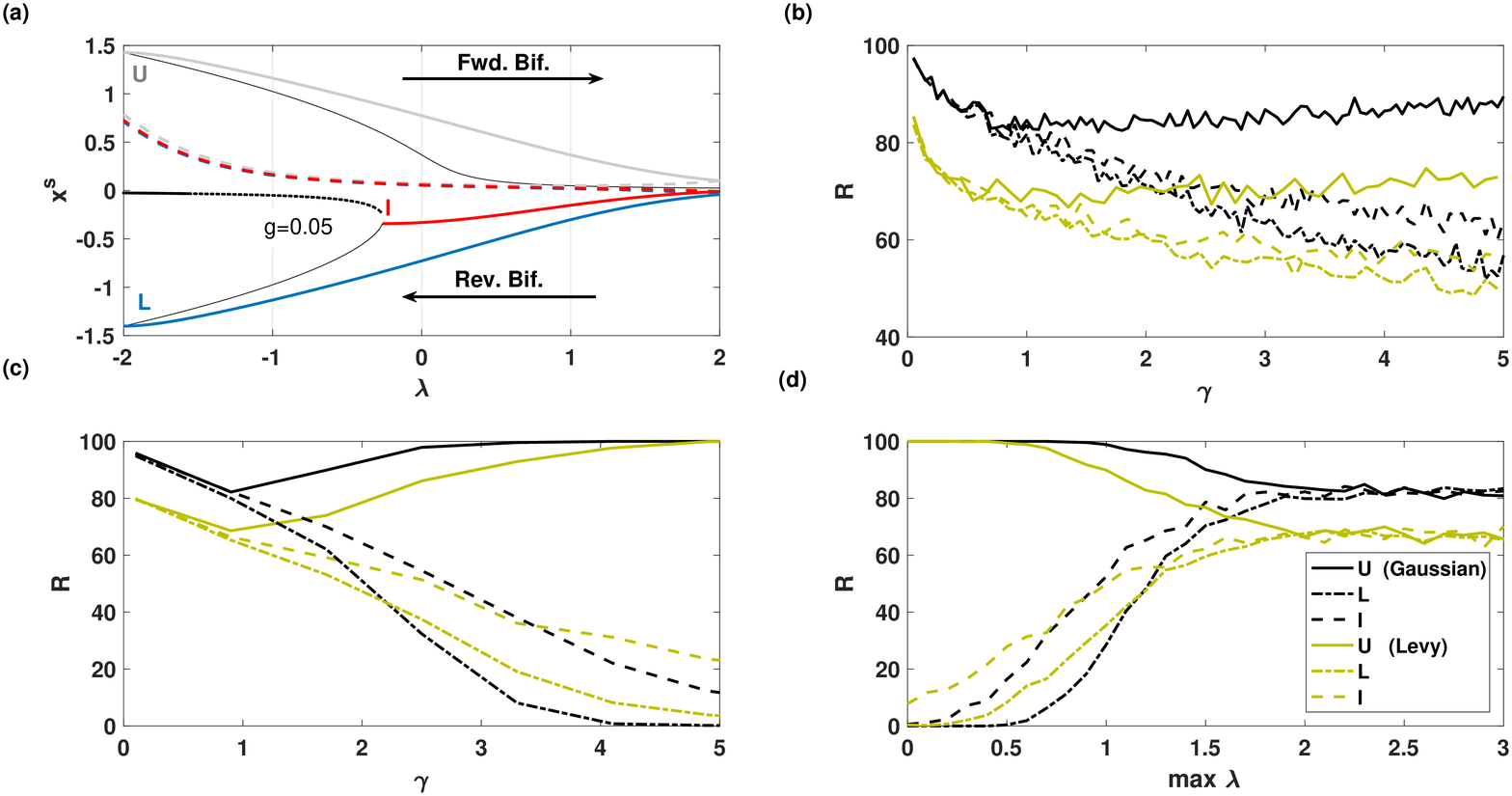}
\caption{Sensitivity of branch selectivity to initial conditions, sweeping speed and amplitude in forward-reverse dynamic bifurcations. (a) Bifurcation diagram indicating sweeping directions, starting conditions and deterministic trajectories. $U$: steady-state at upper branch for $\lambda_{0}=-2$ . $I$: intermediate steady state for $\lambda_{0}=\lambda_{c}=3\left(\frac{g}{2}\right)^{\frac{2}{3}}$. $L$: lower steady-state for $\lambda_{0}=-2$. (b) Selectivity $R$ with $\gamma_{Fwd}=1$ and $\gamma_{Rev}=\gamma$. (c) $\gamma_{Fwd}=\gamma_{Rev}=\gamma$. (d) $\gamma_{Fwd}=\gamma_{Rev}=1$ and maximum amplitude of $\lambda$.  Black lines: Gaussian noise. Green: L\'{e}vy noise, with $\mu = −1.8$, $c = 1/100$ and $\xi^{u} = 5$ (see Eq.~\ref{eq:LevyCDF} and \ref{eq:LevyDensity}). 1000 trajectories were used in the calculation of $R$, the percentage attracted to $x_{+}$. $g=0.05$. $\sigma=0.1$. Fwd: forward segment. Rev: reverse segment.}\label{fig:Fig_FwdRevBifGaussianLevy}
\end{figure*}
% ----------------------
% ----------------------

Typical external signals in biology have complex profiles \cite{Behar2010} and adequate responses to each of the signal characteristics has to occur, to an extent, in the induced expression patterns \cite{Murphy2002,Werner2005}. Previously, we proved that the swicthing delays and the asymmetries in expression patterns induced by external signals can be understood by the simple normal forms represented in Eq.~\ref{eq:PitchFork} \cite{Nene2012,Nene2013} (see also section~\ref{sec:FwdBif}). Yet, as remarked before, the effect of signals on decision making do not push the system in one direction only as they usually return to basal levels; this clearly induces crossing of the critical region in the reverse direction \cite{Werner2005} (see Fig.~\ref{fig:Fig_FwdRevBifGaussianLevy} (a)). Moreover, the nature of signals and networks in biology dictates that the drivers are often compounded \cite{Nicolis2005a} and stochastic \cite{Dattani2016}. A forward-reverse simulation experiment stands, therefore, as a closer representation of the dynamical behaviour of the typical circuitry determining cellular decision making. An interesting contribution to the subject of recurrent bifurcations was also explored through deterministic forcings, although the effects of stochasticity were not approached \cite{Nicolis2003} and the motivation was not the study of biological networks. Here, in order to understand the main ingredients at play in these complex scenarios, we generalize the sweeping process in both directions; the system starts in the bi-stability region, crosses into the monostability region and inverts the movement back to the parameter value it started. This is represented in Fig.~\ref{fig:Fig_FwdRevBifGaussianLevy} (a), which can be reproduced by changing $\lambda(t)$ to $- \lambda (t)$ in Eq.~\ref{eq:PitchFork}.

\subsubsection{Effects of sweeping speed and stochasticity for different initial conditions}

The scenario explored in section~\ref{sec:FwdBif} helps us understand each stage of the experiment represented in Fig.~\ref{fig:Fig_FwdRevBifGaussianLevy} (a): the forward sweeping segment destroys the memory of the initial conditions; the backward segment, studied in section~\ref{sec:FwdBif}, takes the degradation of the initial information encoded in the state of the system at the point of reversal, and tries to recover the position at $t=0$. It should be pointed out that, at a particular sweeping speed $\gamma$, if the sweeping amplitude is large enough, convergence to the upper branch is always present in a deterministic setting (see trajectories in Fig.~\ref{fig:Fig_FwdRevBifGaussianLevy} (a)). The presence of fluctuations ($\xi(t)$ in Eq.~\ref{eq:EqWithNoise}), which represent stochastic processes inherent to each stage of the integrative genetic circuits \cite{Shahrezaei2008,Bowsher2012,Bowsher2013,Bowsher2014,Dattani2016}, hinder the capacity of the system to respond to external signals: here the time-dependent profile of $\lambda$  and $g$. In this sense, both sources contain information that is processed by the normal form. The initial condition constitutes the third source. 

If we start the system at $U$, $L$ or $I$ represented in Fig.~\ref{fig:Fig_FwdRevBifGaussianLevy} (a), it either retains or loses the information contained at the initial instant while responding and processing $\gamma(t)$ and $g$.  During the forward segment the trajectories tend to converge to the only available steady-state solution after crossing the critical region at $-\lambda_{c}$. This convergence is affected by the same parameters as the scenario explored in a previous section where one-way only  sweeps were included. If the sweeping speed is sufficiently low, the system is allowed to converge in the monostability region. Once $\lambda$ is forced back to the starting point, the backward segment, the propensity to be captured by the basin of attraction of the branch favoured by $g$ changes much more if the simulations start at $L$ and $I$. In either, larger sweeping rates reduce the sensitivity of the system to $g$ but increase the likelihood of maintaining memory of the initial condition (see Fig.~\ref{fig:Fig_FwdRevBifGaussianLevy} (c)); lower sweeping rates have the opposite effect. The reasons behind this can also be understood by inspecting the deterministic trajectories in Fig.~\ref{fig:Fig_FwdRevBifGaussianLevy} (a). Since given enough sweeping amplitude convergence to the upper branch is always observed in a deterministic setting, the likelihood of the system converging to the original lower branch is increased only if the differences in the system's relaxation time-scale and that of $\lambda(t)$ are significantly different \cite{Kondepudi1983,Berglund2006}. A similar reasoning holds for trajectories starting at $U$, although the convergence properties after reaching the monostable region are different. In this case, lower sweeping rates secure that the upper bifurcation branch $x_{+}$ is tracked at all times during the forward segment of the forward-backward experiment. If the same sweeping rate is held in the backward segment, the trajectories always track the connected branch and sensitivity to $g$ is secured. In addition, memory of the initial condition is also present. Yet, an interesting feature is verified when $\gamma$ is increased from low values to values above 1. In this region, initially the trend is as expected: higher rates destroy information despite still helping to track the upper branch in the forward sweeping segment. Although the reverse segment is done at the same speed, at this stage it is sufficient to put the mean value among replicated trajectories closer to the unstable boundary which, as was explained above, enhances the propensity to jump across the potential barrier. On the other hand, the chances of remaining in the attractor basin of the upper branch, presuming we start at $U$, increase once again to very high values as we cross the threshold of $\gamma \approx 1$. This stems from not tracking the connected branch and the resulting distance to the stable solution once the monostability region is reached. Not being able to converge fast enough secures reduced branch to branch transitions once the system is reversed and, naturally, an improvement in $R$. This trend also occurs when L\'{e}vy noise is used, although consistently with its typical shape, selectivity is smaller than that achieved with the typical Guassian noise. For the results pertaining to the starting points $L$ and $I$, once again an interesting feature is observed when the heavy-tail distribution is used. Despite an increase in sweeping rate inducing the expected results, the relative magnitude with respect to the results obtained with the Gaussian noise term is inverted.  This can also be attributed to the likelihood of larger positive deviations being more prevalent in the L\'{e}vy distribution, which in combination with the fact that very large sweeping rates trap the system in the starting basin of attraction, improves the relative sensitivity to $g$ (see Fig.~\ref{fig:Fig_FwdRevBifGaussianLevy} (c)). 

Also regarding the importance of critical parameter sweeping speed in the forward and backward segments, Fig.~\ref{fig:Fig_FwdRevBifGaussianLevy} (b) demonstrates that if $\gamma_{Fwd}$ is held at 1 and $\gamma_{Rev}$ is varied, the effects registered before for equal rates is less pronounced, despite the general tendency being the same. Therefore, differences in the sweeping segments may also be a potential mechanism for optimal cellular decision making. This is reminiscent of path-dependent effects recently observed in bio-circuits regulating pattern selection \cite{Palau-Ortin2015}, of expression dynamics behind stress-induced response \cite{Young2013} and, to an extent, of high-dimensional versions of the integrated circuit represented in Fig.~\ref{fig:Fig_BioNet} (a) \cite{Nene2012b}. The combination of time-dependent signals and their shape \cite{Werner2005}, including ascending and descending rates, may have an influence on the probability of reaching certain attractors/cell fates.  

\subsubsection{Varying the elapsed time before system reversal}

When simulating the system according to the same numerical recipe as that presented in the previous section, it is of interest to inspect how the amount of time elapsed before the system is reversed affects the number of paths attracted to each attractor.  For maximum values of $\lambda$ below 1, $100\%$ of the paths converge to the positive steady state if the starting point is $U$ and the noise model is Gaussian (see Fig.~\ref{fig:Fig_FwdRevBifGaussianLevy} (a) and (d)). This slowly decreases as maximum amplitudes of $\lambda$ are gradually increased to 3. Attaining larger values of $\lambda$ before reversal allows for the convergence of the system to the solutions represented by the upper branch, which is favoured by the constant external asymmetry. This is fundamentally important because although the drift rate is approximately $g$, the relaxation to the equilibrium in the monostability region is quite slow for values of $\lambda$ not far from $\lambda_{c}$; for example, at  $\lambda=0$, the position of the steady state is approximately $(g(\lambda=0))^{1/3}$, which makes the relaxation time $~(g(\lambda=0))^{-2/3}$ \cite{Kondepudi1983}. Therefore, larger maximum amplitudes increase the convergence rate. On the other hand, proximity to the stable positive branch at large $\lambda$ values implies proximity to the unstable branch if the sweeping speed is sufficiently high once the critical parameter is reversed (see also section~\ref{sec:FwdBif}). This, in turn, affects the capacity of the system to retain information of the starting condition due to the importance of the sweeping rate in enhancing the likelihood of escape, especially in fluctuation distributions with larger jumps (see also Fig.~\ref{fig:Fig_FwdBifGaussianLevy}). 

For lower starting positions ($L$ or $I$ in Fig.~\ref{fig:Fig_FwdRevBifGaussianLevy} (a)), the reverse scenario is observed (Fig.~\ref{fig:Fig_FwdRevBifGaussianLevy} (d)). As the forward sweeping maximum $\lambda$ amplitude is increased the more efficient the asymmetry is; towards larger values all starting positions attain roughly a selectivity of $80\%$ and $60\%$ for Gaussian and L\'{e}vy noises, respectively. A similar reasoning as that presented above is valid here. Yet, the tendency observed is that larger amplitudes lead to an improvement of the effectiveness of $g$ as a state selector in the face of fluctuations.

The effect of the distribution of fluctuations is once again verified, especially for starting positions $L$ and $I$ (see Fig.~\ref{fig:Fig_FwdRevBifGaussianLevy} (a)): the L\'{e}vy distribution leads to lower selectivities for larger sweeping amplitudes, i.e when proximity of sample paths to the unstable state is more probable. On the other hand, the larger asymmetry for positive jumps in $\xi$ (see Eq.~\ref{eq:EqWithNoise}) when a L\'{e}vy noise term is used works to increase the relative $R$ if lower sweeping amplitudes are imposed, i.e. when distances to the unstable branch are such that the Gaussian noise is not as successful in eliciting jumps into the basing of attraction of the selected branch. Despite the fact that the L\'{e}vy distribution has a long tail towards positive jumps, which effectively secured an overwhelming bias towards the positive branch for large $\sigma$'s when one-way only dynamic bifurcations were tested (section~\ref{sec:StochFluct}), the results plotted in Fig.~\ref{fig:Fig_FwdRevBifGaussianLevy} (d) were derived with $\sigma=0.1$. At this amplitude the imbalance towards the upper branch is still relatively minor (see Fig.\ref{fig:Fig_FwdBifGaussianLevy} (b) and Eqs.\ref{eq:LevyCDF} and \ref{eq:LevyDensity}) and its success in increasing selectivity is only secured in conjunction with the other ingredient tested in this section. This synergy is, in many ways, similar to that observed in Fig.~\ref{fig:Fig_FwdRevBifGaussianLevy} (c), where lower sweeping rates exert a similar action to that of larger sweeping amplitudes. We must also add that the order of the respective selectivity curves is consistent with the distance of the initial conditions to the steady-state favoured by $g$.

\section{Discussion and further work}

Several important contributions stemming from non-equilibrium physics have been applied to the problem of information processing in bio-circuits \cite{Berg2008,Ge2009,Kobayashi2011}. The equally rich field of open-systems \cite{Cross1993,Hobbs2013,Perryman2014} and dynamic bifurcations \cite{Erneux1991,Berglund2006}, which deals with equivalent problems, has been less utilized in the interpretation of biological intracellular phenomena. The mechanism of Speed-dependent Cellular Decision Making, initially proposed in \cite{Nene2012}, and further advanced in \cite{Nene2013}, contributes to the expansion of this field in biology. Here, we developed the framework further by testing the ability of dynamically bifurcating systems to retain memory of initial conditions in the face of forward-reverse sweeping through critical regions and heavy-tailed noise distributions. This systematic investigation is clearly in line with the effects of complex signals that dictate encoded evolutionary responses to environmental pressures. Moreover, asymmetric heavy-tail L\'{e}vy distributions have recently been proposed as viable alternatives that naturally incorporate large deviations even at small noise amplitudes. In order to analyse clearly all of the elements underlying rate-dependent phenomena in fluctuating systems, we resorted once more to the paradigmatic bi-stable potential problem undergoing a supercritical Pitchfork bifurcation. This simple system was proven to exhibit similar characteristics to representative intra-cellular circuits and constituted a simple approach allowing for thorough computational tests.

Overall, sweeping through the critical region at different rates has different effects on correct branch identification when we start at different initial conditions in forward-backwards dynamic bifurcations. Whereas a slow passage through the critical region may help to process the information carried by an external asymmetry and, additionally, a gradual increase in sweeping rates degrades this sensitivity, this is only strictly true in forward bifurcations from a monostability to a bi-stability region. In forward-backwards sweeps, if the system starts at the branch favoured by the external signal, monotonicity of state selectivity with sweeping rate is not observed. A region in the vicinity of sweeping rates close to 1 hinders both the maintenance of memory of initial conditions and the effect of external asymmetries. For initial conditions not in the attractor basins of the state favoured by the external signal the gradual tendency with sweeping rates is similar to that observed in previous studies.

Moreover, it is the combination of sweeping speed and noise distribution that allows for a robust memory of starting conditions. Heavy-tail distributions can destroy all information encoded in an external signal if the sweeping speed is not adapted to the starting point and the amplitude of maximal deviation in a forward-backwards dynamic bifurcation. Additional tests on other bifurcations that can explain decision making in biology \cite{Huang2007,Enver2009,Wang2011} should also reveal specific balances between bifurcation type, type of external signal and unexpected rate-dependent effects contributing to correct information processing in the face of large fluctuations.

Regarding the rate-dependent forward-reverse dynamic bifurcation as a mechanism for decision making, we observed that, if sweeping speeds and amplitudes are sufficiently low and high, respectively, this is fundamentally a more efficient strategy for processing signals than attractor to attractor transitions over potential barriers, if the system is initially at a "sub-optimal" position. The latter decision making mechanism has been accepted in the literature as a strategy used by bio-circuitry under uncertainty. The field of dynamic bifurcations has been less explored as a tool. Yet, as was proven in previous work and in the present paper it is a viable alternative that should be explored in real cellular networks.

\section{Acknowledgements}
AZ acknowledges support from the Russian Federation grant 14.Y26.31.0026.

\bibliographystyle{apsrev4-1}
\bibliography{RevBif2018}

\end{document}